\documentclass{PoS}

\title{
\vspace{-20mm}
 \hspace{120mm}
 {\rm \normalsize KEK-TH-1880}
 \\[15mm]
Understanding the problem with logarithmic singularities 
in the complex Langevin method}

\ShortTitle{Understanding the problem with logarithmic singularities...}

\author{Jun Nishimura\\
        KEK Theory Center, High Energy Accelerator Research Organization, Tsukuba 305-0801, Japan\\ 
        Department of Particle and Nuclear Physics, School of High Energy Accelerator Science, 
        Graduate University for Advanced Studies (SOKENDAI), Tsukuba 305-0801, Japan\\
        E-mail: \email{jnishi@post.kek.jp}}

\author{\speaker{Shinji Shimasaki}
  \\
        KEK Theory Center, High Energy Accelerator Research Organization, Tsukuba 305-0801, Japan\\
        Research and Education Center for Natural Sciences, Keio University, Hiyoshi 4-1-1, Yokohama, 
        Kanagawa 223-8521, Japan \\ 
        E-mail: \email{simasaki@post.kek.jp}}

\abstract{In recent years, there has been remarkable progress in theoretical justification 
of the complex Langevin method, which is a promising method for evading the sign problem in the path integral 
with a complex weight. There still remains, however, an issue concerning occasional failure of this method 
in the case where the action involves logarithmic singularities such as the one appearing from the fermion 
determinant in finite density QCD.
In this talk, we point out that this failure is due to the breakdown of the relation 
between the complex weight which satisfies the Fokker-Planck equation and the probability distribution 
generated by the stochastic process. In fact, this kind of failure can occur in general 
when the stochastic process involves a singular drift term. We show, however, in simple examples, 
that there exists a parameter region in which the method works although the standard reweighting method 
is hardly applicable.}

\FullConference{The 33rd International Symposium on Lattice Field Theory\\
		14 -18 July 2015\\
		Kobe International Conference Center, Kobe, Japan*}

\begin{document}

\section{Introduction}

Monte Carlo method is a powerful tool to study quantum field theories nonperturbatively.
However, this method is
based on the probability interpretation of the Boltzmann weight $e^{-S}$.
Therefore, for theories with a complex action such as finite density QCD, 
real time dynamics of quantum mechanics, supersymmetric gauge theories and so on,
we cannot use Monte Carlo methods in a straightforward manner.

The complex Langevin method is a method 
which can potentially solve this complex action problem \cite{Parisi,Klauder}.
This method is a naive extension of 
the ordinary stochastic quantization \cite{ParisiWu} for a real action 
to the case of a complex action,
in which real dynamical variables obeying the Langevin equation are complexified.
The expectation values in the original path integral 
are computed by taking the statistical average 
of the corresponding quantities over the stochastic process. 
(See ref.~\cite{Damgaard} for a review.)
However, this method is not yet established as a solution
to the complex action problem
due to a well-known issue that it sometimes converges to a wrong result.

Recently, remarkable progress has been made in understanding the necessary condition for 
the CLM to give correct results \cite{Aarts1101}.
The crucial point for the CLM to work is that the probability distribution defined 
from the stochastic process can be converted to the complex weight 
in the original path integral.
From this point of view, the breakdown of the above relation
has been identified as the cause of 
the failure that occurs when the probability distribution has a slow fall-off
in the imaginary direction.


It was also reported that the method can fail
when the action involves 
logarithmic singularities in the action \cite{Mollg1309,Mollg1412,Green1406}.
A typical example is QCD at finite density,
where the effective action involves the logarithm of the fermion determinant,
which becomes complex for nonzero chemical potential.
It was shown in a simplified model that 
a naive implementation of the CLM can give
wrong results close to those for the phase quenched model.
This happens when the phase of the fermion determinant 
frequently rotates, and this observation led the authors to
speculate that this failure has something to do with 
the ambiguity of the logarithm associated with the branch cut.

In this talk, we clarify the cause of the above failure 
from the viewpoint of the crucial relation 
between the probability distribution and the complex weight \cite{Nishi1504}.
We show that the failure is not 
caused by the ambiguity of the logarithm
but rather by the singularities involved in the drift term,
which invalidate the crucial relation.

\section{A simple example}
\subsection{Complex Langevin method}

As a simple example, let us consider the path integral with a real variable $x$ given by
\begin{equation}
  Z= \int_{-\infty}^{\infty} dx \, w(x) \ , 
\qquad w(x)=(x+i\alpha)^p e^{-x^2/2} \ , 
\label{Z}
\end{equation}
where $\alpha$ and $p$ are real parameters.
When $p\neq 0$ and $\alpha\neq 0$, the weight $w(x)$ becomes complex and 
gives rise to the complex action problem.

Now we apply the CLM to this system. First, we define the drift term as
\begin{eqnarray}
v(x)=w(x)^{-1}\frac{\partial w(x)}{\partial x}=\frac{p}{x+i\alpha}-x \ .
\end{eqnarray}
Then, we have to complexify the real variable $x$ as
$x\rightarrow z=x+iy$.
The action is given in terms of the complexified variable $z$ as
\begin{equation}
S(z)=\frac{1}{2}\, z^2 - p\log(z+i\alpha) \ . 
\label{S}
\end{equation}
Clearly, $S(z)$ involves a logarithmic singularity at $z=-i\alpha$ for $p\neq 0$,
which causes the ambiguity associated with the branch cut. 
However, this is not an issue since 
the Langevin process 
corresponding to the path integral (\ref{Z})
can be formulated by using the drift term $v(z)$, 
and the action (\ref{S}) is not necessary.
In fact, all we need is the single-valuedness of the drift term $v(z)$ 
as a function of the complexified variable $z$,
and the single-valuedness of the complex weight $w(x)$ 
as a function of the original real variable $x$.

The complex Langevin equation is given by
\begin{equation}
\frac{dz}{dt}=v(z)+\eta(t)=\frac{p}{z+i\alpha}-z+\eta(t) \ , 
\label{Langevin eq}
\end{equation}
where $\eta(t)$ is a real Gaussian noise 
normalized as $\langle \eta(t)\eta(t')\rangle_\eta =2\delta(t-t')$.
We denote the solution of (\ref{Langevin eq}) 
as $z^{(\eta)}(t)=x^{(\eta)}(t)+iy^{(\eta)}(t)$.
The probability distribution at time $t$ is defined by
$P(x,y;t)=\langle \delta(x-x^{(\eta)}(t))\delta(y-y^{(\eta)}(t))\rangle_\eta$ and
its time evolution is given by the Fokker-Planck-like equation
\begin{eqnarray}
\frac{\partial}{\partial t}P(x,y;t)
&=&L^\top P(x,y;t) \nonumber\\
&=&\frac{\partial}{\partial x}\left\{-\left(\mathrm{Re}v\right)|_{z=x+iy}+\frac{\partial}{\partial x}\right\} P(x,y;t)
 +\frac{\partial}{\partial y}\left\{-\left(\mathrm{Im}v\right)|_{z=x+iy}\right\} 
P(x,y;t) \ .
\end{eqnarray}

\subsection{Condition for correct convergence}

The justification of the CLM follows from the relation 
\begin{equation}
\int dxdy \, \mathcal O(x+iy)P(x,y;t) = \int dx \, \mathcal O (x) \rho(x;t) 
\label{Prho}
\end{equation}
between the probability distribution $P(x,y;t)$ and the complex weight $\rho(x;t)$,
where $\mathcal O(x)$ is an observable 
which admits the holomorphic extension to $\mathcal O(x+iy)$.
Here the complex weight $\rho(x;t)$ obeys the Fokker-Planck equation
\begin{equation}
\frac{\partial}{\partial t} \rho(x;t) = L_0^\top \rho(x,t) \ , \quad
L_0^\top = 
\frac{\partial}{\partial x}\left(\frac{\partial}{\partial x}-v(x)\right) \ . \label{rhozt}
\end{equation}
Note that the complex weight $w(x)$ appearing
in the original path integral (\ref{Z})
is a time-independent solution of (\ref{rhozt}).
In ref.~\cite{Aarts1101}, the relation (\ref{Prho}) 
is derived by showing the following identities
\begin{eqnarray}
\int dxdy \, \mathcal O(x+iy)P(x,y;t) 
&=& \int dxdy \, \mathcal O(x+iy;t)P(x,y;0)\ , \label{PO} \\
\int dx \, \mathcal O(x)\rho(x;t) &=& 
\int dx \, \mathcal O(x;t) \rho(x;0) \ , \label{rhoO}
\end{eqnarray}
where $\mathcal O(x+iy;t)$ is defined by
\begin{equation}
\frac{\partial}{\partial t}\mathcal O(z;t) = \tilde L \mathcal O(z;t)\ , \quad
\tilde L=
\left(\frac{\partial}{\partial z}+v(z)\right)\frac{\partial}{\partial z}\ , \label{Ozt}
\end{equation}
and $\mathcal O(z;0)=\mathcal O(z)$. At arbitrary $t$, $\mathcal O(z;t)$ is shown to be holomorphic.
$\mathcal O(x;t)$ is a restriction of $\mathcal O(x+iy;t)$ to $y=0$ and obeys 
\begin{equation}
\frac{\partial}{\partial t}\mathcal O(x;t) = L_0 \mathcal O(x;t) \ , \quad
L_0=\left(\frac{\partial}{\partial x}+v(x)\right)\frac{\partial}{\partial x}\ . 
\label{Oxt}
\end{equation}
We assume $P(x,y;0)=\rho(x)\delta(y)$ 
as the initial condition, which makes
the right-hand sides of (\ref{PO}) and (\ref{rhoO}) equal to each other.
To show (\ref{PO}), we first define
\begin{equation}
F(t,\tau)=\int dxdy \, \mathcal O(x+iy;\tau) \, P(x,y;t-\tau) \ , 
\label{F}
\end{equation}
which interpolates both sides of (\ref{PO}) with $0\leq \tau \leq t$.
Let us consider the $\tau$-derivative of (\ref{F})
\begin{equation}
\frac{d}{d\tau}F(t,\tau)
=\int dxdy \, \tilde L \mathcal O(x+iy;\tau)P(x,y;t-\tau)
  -\int dxdy \, \mathcal O(x+iy;\tau)\, L^\top P(x,y;t-\tau) \ .
\end{equation}
This quantity can be shown to vanish 
by using the integration by parts and the holomorphy of $\mathcal O(x+iy;t)$.
Therefore, $F(t,\tau)$ is $\tau$-independent, and hence (\ref{PO}) follows.
However, in order to perform the partial integration,
the boundary terms should be neglected,
which requires that the probability distribution 
has a fast fall-off at $|x|,|y|\rightarrow \infty$.
One can also show the relation (\ref{rhoO}) in a similar way. 
In this case, the boundary terms 
vanish due to the effects of the action.

\subsection{Boundary terms due to singularities}

The present example (\ref{Z}) does not 
suffer from the problem due to the boundary terms 
at $|x|,|y|\rightarrow \infty$ 
since the probability distribution has a fast fall-off
due to the drift term, which behaves as $\sim -z$.
However, the drift term involves a singularity at $z=-i\alpha$, 
which causes the problem of boundary terms
in performing the integration by parts \cite{Nishi1504}.
In order for the boundary terms at the singularity to be neglected, 
the following quantities have to be finite for arbitrary $t$ and $\tau$.
\begin{eqnarray}
&&\lim_{x\rightarrow 0}\left[x\int \frac{\mathcal O(z;\tau)}{|z+i\alpha|^2}
\, P(x,y;t-\tau)dy\right] \ , \\
&&\lim_{y\rightarrow -\alpha}
\left[(y+\alpha)\int \frac{\mathcal O(z;\tau)}{|z+i\alpha|^2}
\, P(x,y;t-\tau)dx\right] \ .
\end{eqnarray}
Note that $\mathcal O(z;\tau)$ defined as a solution to (\ref{Ozt})
is highly singular at $z=-i\alpha$ 
since $\tilde L$ on the right-hand side of (\ref{Ozt}) is singular.
Thus, in order to satisfy the above conditions,
the probability distribution should practically vanish near the singularity.

\section{Numerical results}

In this section, we demonstrate our assertion in the previous section
by showing the results of the CLM in a few examples \cite{Nishi1504}.

\subsection{One-variable case}

\begin{figure}[h]
  \begin{minipage}{1.0\hsize}
    \centering  \includegraphics[width=6.5cm]{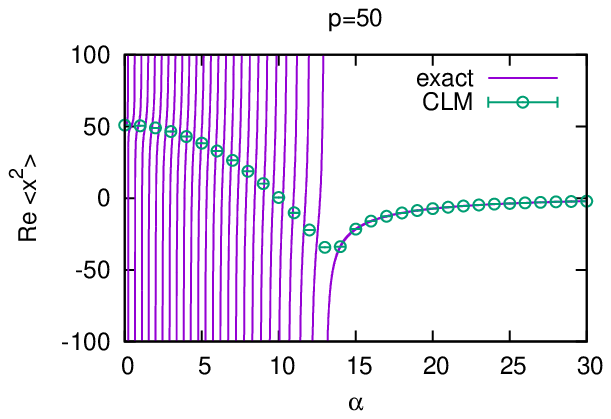}
    \hspace{0cm}\vspace{3mm}
    \centering
    \caption{The result of the CLM for the case (\protect\ref{Z}) with $p=50$.
      The real part of $\langle x^2\rangle$ is plotted against $\alpha$.
The solid line represents the exact results.}
    \label{p50}
  \end{minipage}\\
  \begin{minipage}{0.5\hsize}
    \centering  \includegraphics[width=6.5cm]{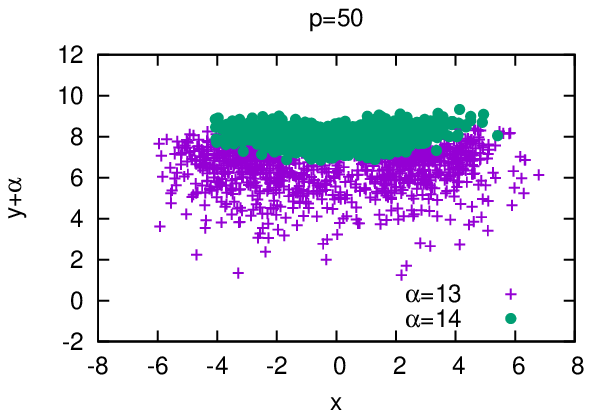}
    \hspace{0cm}
    \centering\caption{The scattered plot obtained by the CLM for 
the case (\protect\ref{Z}) with $p=50$. 
      The results for $\alpha=13$ and $\alpha=14$ are plotted.}
   \label{p50scatter}
  \end{minipage}
  \hspace{5mm}
  \begin{minipage}{0.5\hsize}
    \centering \includegraphics[width=6.5cm]{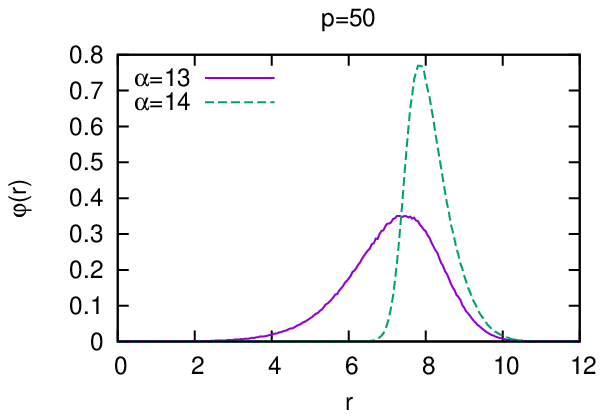}
    \hspace{0cm}
    \centering\caption{The radial distribution obtained by the CLM 
for the case (\protect\ref{Z}) with $p=50$. 
      The results for $\alpha=13$ and $\alpha=14$ are plotted.}
   \label{p50radial}
  \end{minipage}
\end{figure}

In Figure~\ref{p50}, 
we show the results obtained by the CLM
for the one-variable example (\ref{Z}) with $p=50$.
We observe that our results deviate from the exact values at $\alpha\lesssim 13$.
This failure is due to the singular-drift problem 
discussed in the previous section.
Indeed, one can see from Figure~\ref{p50scatter} that,
while the distribution obtained for $\alpha=14$ is 
away from the singularity,
it comes closer to the singularity for $\alpha=13$.
The situation is clearer in Figure~\ref{p50radial}, 
which shows the radial distribution defined by
\begin{equation}
\varphi(r)=\frac{1}{2\pi r}\int dxdy \, P(x,y;\infty)
\, \delta(\sqrt{x^2+(y+\alpha)^2}-r) 
\end{equation}
corresponding to Figure~\ref{p50scatter}. 
Note that the phase of $(z+i\alpha)^{50}$ rotates frequently in the Langevin process
even for $\alpha\gtrsim 14$, where the CLM works well. 
Thus, the failure of the CLM at $\alpha\lesssim 13$  should be attributed not
to the ambiguity of the logarithmic term
but rather to the singularity of the drift term.
This point becomes more manifest in the next example.

\subsection{Non-logarithmic case}

\begin{figure}
  \begin{minipage}{0.5\hsize}
    \centering  \includegraphics[width=6.5cm]{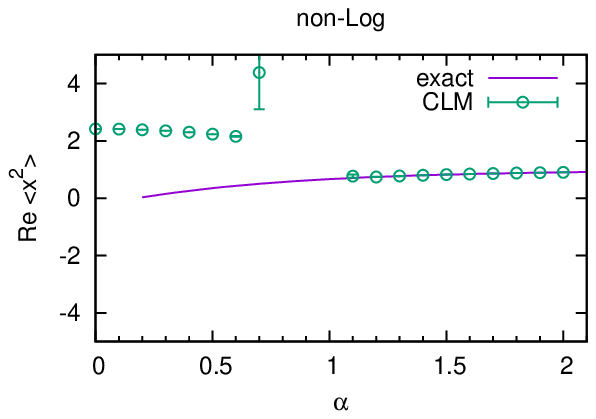}
    \hspace{0cm}
    \centering
    \caption{The result of the CLM 
for the case (\protect\ref{nonlog}) with a non-logarithmic singularity. 
The real part of $\langle x^2\rangle$ is plotted against $\alpha$.}
    \label{nonlogfig}
  \end{minipage}
  \hspace{5mm}
  \begin{minipage}{0.5\hsize}
    \centering \includegraphics[width=6.5cm]{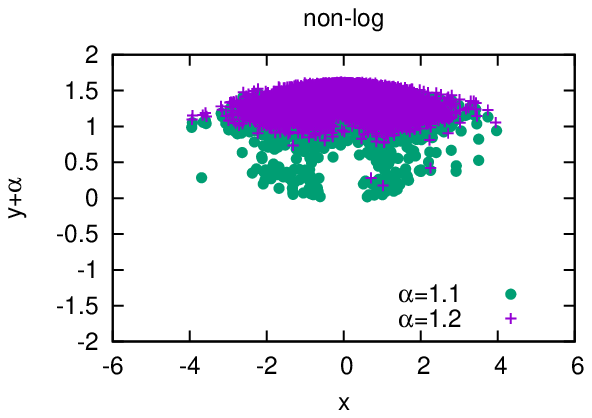}
    \hspace{0cm}
    \centering\caption{The scattered plot obtained by the CLM 
for the case (\protect\ref{nonlog}) with a non-logarithmic singularity. 
      The results for $\alpha=1.1$ and $\alpha=1.2$ are plotted.}
    \label{nonlogfigscatter}
  \end{minipage}
\end{figure}

Here we discuss an example 
in which the action involves a non-logarithmic singular term;
namely we consider the action
\begin{eqnarray}
S(x)=(x+i\alpha)^{-2}+\frac{1}{2}\,  x^2 \ . \label{nonlog}
\end{eqnarray}
Upon complexification $x\rightarrow z=x+iy$, 
the action (\ref{nonlog}) has no ambiguity associated with the branch cut,
but the drift term $v(z)=2(z+i\alpha)^{-3}+z$ has a singularity at $z=-i\alpha$.
The results of the CLM for this system are shown 
in Figure~\ref{nonlogfig} and Figure~\ref{nonlogfigscatter}.
We find from Figure~\ref{nonlogfig} that 
the CLM works well at $\alpha\gtrsim 1.2$, but it fails at smaller $\alpha$.
It turns out from Figure~\ref{nonlogfigscatter} that in the case of small $\alpha$,
the distribution comes close to the singularity of the drift term.
Thus, the failure of the CLM should be attributed to the singular-drift problem. 

\subsection{Two-variable case}

Our arguments can be extended to the multi-variable case.
To demonstrate this, let us consider a system
with two real variables $x_1$ and $x_2$ given by
\begin{equation}
Z=\int dx_1dx_2 \, w(x_1,x_2)\ , \quad
w(x_1,x_2)= (x_1+i x_2)^p \, e^{-\frac{1}{2}(x_1)^2-\frac{1}{2}(x_2-\alpha)^2} \ ,
\label{2d}
\end{equation}
where $p$ is a positive integer and $\alpha$ is a real parameter.
In the CLM, $x_1$ and $x_2$ are complexified to $z_1$ and $z_2$,
respectively, 
and the drift term becomes singular on the surface defined by $z_1+iz_2=0$.

\begin{figure}
  \begin{minipage}{0.5\hsize}
    \centering \includegraphics[width=6.5cm]{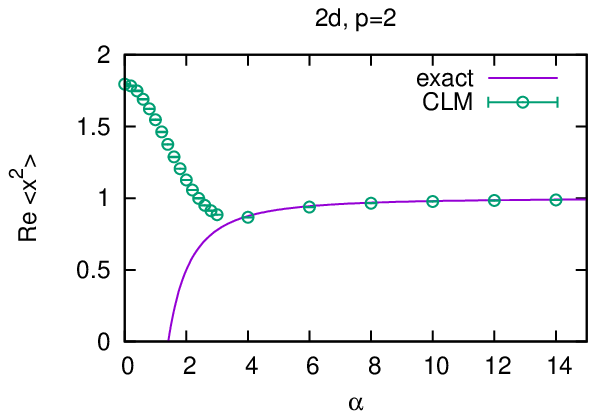}
    \hspace{0cm}
    \centering
    \caption{The result of the CLM 
      for the two-variable case (\protect\ref{2d}) with $p=2$.
      The real part of $\langle x_1^2\rangle$ is plotted against $\alpha$.}
    \label{2dp2}
  \end{minipage}
  \hspace{5mm}
  \begin{minipage}{0.5\hsize}
    \centering \includegraphics[width=6.5cm]{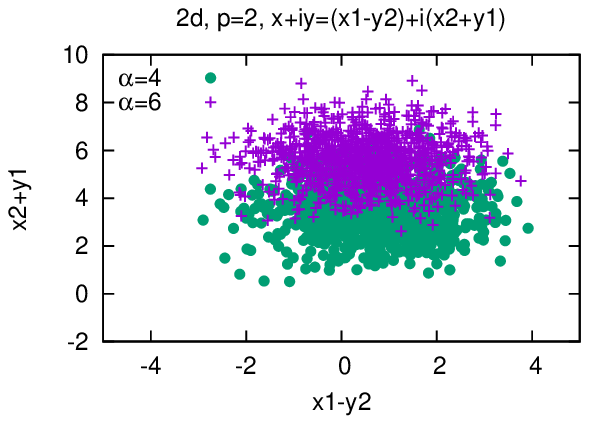}
    \hspace{0cm}
    \centering\caption{The scattered plot obtained by the CLM
 for the two-variable case (\protect\ref{2d}) with $p=2$.
      The results for $\alpha=4$ and $\alpha=6$ are plotted.}
    \label{2dp2scatter}
  \end{minipage}
\end{figure}

The results of the CLM
for this case with $p=2$ are given in Figure~\ref{2dp2} 
and Figure~\ref{2dp2scatter}.
We find that the CLM works at $\alpha\gtrsim 6$,
and from Figure~\ref{2dp2scatter}
we find that the distribution comes close to the 
singularity for smaller $\alpha$.

\section{Summary and Discussion}

We have investigated the issue of the CLM 
that arises when the action involves logarithmic singularities.
The failure of the CLM in some parameter region
was thought to have something to do with 
the multi-valuedness of the action for the complexified variables. 
We pointed out, however, that this cannot be considered as a cause of the problem
since one can formulate the CLM without referring to the action.
We then discussed this issue based on the key relation (\ref{Prho})
between the probability distribution and the complex weight. 
We have shown that the relation can be violated 
unless the probability distribution is practically zero around the singularities
since in such a case the boundary terms 
appearing in the integration by parts used to show (\ref{Prho}) cannot be neglected.
We provided some evidence supporting our assertion
by applying the CLM to some simple examples.

Based on our new understanding, we may be able to develop a method 
to avoid the problem.
One possibility is to extend the idea of the gauge cooling \cite{Seiler1211}. 
This technique was originally developed
for the purpose of solving the problem caused by the boundary terms at infinity.
Since the cause of the singular-drift problem also resides 
in the boundary terms that appear in the integration by parts,
we should be able to apply the same technique.

The singular-drift problem is anticipated to occur in finite density QCD
at low temperature with small quark mass,
where small eigenvalues of the Dirac operator tend to be appear.
Indeed the convergence to wrong results has been observed
in the small mass regime 
of the chiral random matrix theory \cite{Mollg1309,Mollg1412},
which is a simplified model of finite density QCD at zero temperature.
In ref.~\cite{Nagata}, we apply a new type of gauge cooling to this model
and show that one can obtain correct results even in the small mass regime. 
We expect that the same technique works also in QCD at finite density.

\end{document}